\documentclass[reprint,amsmath,amssymb,aps,prl]{revtex4-2}

\usepackage{graphicx}
\usepackage{dcolumn}
\usepackage{bm}
\usepackage{multirow}

\begin{document}

\preprint{APS/123-QED}

\title{Core Electron Binding Energies in Solids from Periodic All-Electron $\Delta$-Self-Consistent-Field Calculations}
\thanks{A footnote to the article title}%

\author{J. Matthias Kahk}
 \affiliation{Department of Materials, Imperial College London, South Kensington, London SW7 2AZ, United Kingdom}
 \affiliation{Institute of Physics, University of Tartu, W. Ostwaldi 1, 50411 Tartu, Estonia}
 
\author{Georg S. Michelitsch}
 \affiliation{Chair for Theoretical Chemistry, Technische Universität München, Lichtenbergstr. 4, D-85747 Garching, Germany}
 
\author{Reinhard J. Maurer}
 \affiliation{Chair for Theoretical Chemistry, Technische Universität München, Lichtenbergstr. 4, D-85747 Garching, Germany}
 \affiliation{Department of Chemistry, University of Warwick, Gibbet Hill Rd, Coventry, CV4 7AL, United Kingdom} 
 
\author{Karsten Reuter}
 \affiliation{Chair for Theoretical Chemistry, Technische Universität München, Lichtenbergstr. 4, D-85747 Garching, Germany}

\author{Johannes Lischner}
 \email{j.lischner@imperial.ac.uk}
\affiliation{Department of Physics and Department of Materials, and the Thomas Young Centre for Theory and Simulation of Materials, Imperial College London, London SW7 2AZ, United Kingdom}

\date{\today}

\begin{abstract}
Theoretical calculations of core electron binding energies are important for aiding the interpretation of experimental core level photoelectron spectra. In previous work, the $\Delta$-Self-Consistent-Field ($\Delta$-SCF) method based on density functional theory has been shown to yield highly accurate 1s and 2p binding energies in free molecules. However, most experimental work is concerned with solids, not gases. In this study, we demonstrate the application of the all-electron $\Delta$-SCF method to periodic systems. A consideration of the experimentally accessible points of reference leads to the definition of a core electron binding energy in a solid as the difference between the total energies of two $N-1$ electron systems: one with an explicit, localized core hole, and one with an electron removed from the highest occupied state. The calculation of each of these quantities is addressed. In addition, the analogy between a localized core hole and a charged defect in a solid is highlighted, and the extrapolation of calculated core electron binding energies to the infinite supercell limit is discussed. It is found that the extrapolated values of the core electron binding energies from periodic $\Delta$-SCF calculations agree well with experimental results for both metallic and insulating systems, with a mean absolute error of 0.24 eV for the 15 core levels considered in this study.
\end{abstract}

\maketitle


\emph{Introduction}.---Core electron binding energies are the central concept of core level X-ray Photoelectron Spectroscopy (XPS), which is one of the most widely used experimental techniques in studies of the surfaces of materials. Core level XPS takes advantage of the fact the the energy required to remove a core electron from a particular atom depends on the chemical environment of that atom, to yield information about the chemical environments that are present in a given sample. However, the successful interpretation of a core level photoelectron spectrum is reliant on the availability of good quality reference data. As experimentalists strive to study increasingly complex systems, obtaining such data via experimental means only is becoming ever more challenging.

Theoretical modelling can assist the interpretation of core level XPS spectra by providing an independent means for determining core electron binding energies. Already in 1965, Bagus demonstrated that the $\Delta$-Self-Consistent-Field ($\Delta$SCF) method based on Hartree Fock theory can provide relatively accurate core electron binding energies in isolated atoms \cite{bagus_self-consistent-field_1965}. In recent years, there has been renewed interest in the development of accurate and affordable computational methods for modelling core level photoemission \cite{ozaki_absolute_2017,kahk_accurate_2019,kahk_core_2018,regoutz_combined_2020,golze_accurate_2020,golze_core-level_2018,zhu_all-electron_2021,aoki_accurate_2018,walter_offset-corrected_2016,ljungberg_implementation_2011,lembinen_calculation_2020,liu_benchmark_2019,garcia-gil_calculation_2012,susi_calculation_2015,hait_highly_2020,besley_modeling_2021,vines_prediction_2018,zheng_performance_2019,klein_nuts_2021}. Considerable progress has been made in calculations of free molecules - modern $\Delta$SCF methods based on density functional theory (DFT) using a meta-generalized-gradient-approximation (meta-GGA) functional have been shown to yield highly accurate absolute core electron binding energies, when compared against gas phase measurements.\cite{pueyo_bellafont_performance_2016,kahk_accurate_2019,hait_highly_2020} For example, provided that relativistic effects are properly accounted for, $\Delta$SCF with the SCAN functional yields a mean absolute error (MAE) of just 0.16 eV for a dataset of 103 1s and 2p core electron binding energies \cite{kahk_accurate_2019}. It has also been shown that many-body perturbation theory in the form of full-frequency eigenvalue self-consistent GW calculations can yield accurate core electron binding energies in free molecules \cite{golze_accurate_2020,golze_core-level_2018}. However, most experimental work is concerned with solids, not gases. 

The theoretical prediction of core electron binding energies in solids has proven to be much more challenging. Previous studies that have tackled this issue have either considered only a very small number of systems (\cite{kahk_accurate_2019}), or reported significantly larger errors compared to experiment (\cite{ozaki_absolute_2017,aoki_accurate_2018,walter_offset-corrected_2016,zhu_all-electron_2021}). Therefore, there is still a need for an affordable and accurate method for predicting core electron binding energies in solids. This method should be applicable to both metallic and gapped systems, and to be of greatest use, it should allow for comparisons of core electron binding energies between different samples, not just binding energy shifts within the same sample. The ability to predict binding energies on an absolute energy scale is therefore highly desirable.

The direct application of the DFT-$\Delta$SCF method to periodic systems is faced with some complications. In a conventional $\Delta$SCF calculation, the core electron binding energy is obtained as the total energy difference between systems with different numbers of electrons: the $N$ electron ground state, and the $N-1$ electron final state with an explicit core hole. This makes the problem of determining a core electron binding energy in a periodic solid analogous to the problem of determining the formation energy of a charged defect. As such, the same issues that arise in calculations of charged defects, e.g. defining the chemical potential of the removed electron, eliminating the divergence of the Hartree potential in periodic calculations of charged systems, and minimizing the spurious interactions between periodic images of the defect, are also relevant for modelling core level XPS. Several approaches have been proposed to tackle or circumvent these issues, including the use of cluster models instead of periodic models of solids \cite{kahk_accurate_2019}, promoting a core electron to the lowest energy empty state instead of removing it from the system \cite{pehlke_evidence_1993,olovsson_first_2010}, or using the exact Coulomb cutoff method to remove the spurious interactions between periodic images of the core hole \cite{ozaki_absolute_2017}.

In contrast, in this manuscript we demonstrate that absolute core electron binding energies in periodic solids in fact reduce to total energy differences between systems with the same number of electrons, as long as an appropriate point of reference for the absolute energy scale is chosen. Calculations of charged systems still need to be performed, but the simplest possible remedy of introducing a uniform background charge is found to be adequate. We also address the localization of a core hole in a periodic all-electron DFT calculation, and show that the total energy of the core hole state can be obtained from a calculation where the only constraint that is applied is the non-Aufbau-principle occupation of the Kohn-Sham eigenstates. The proposed method is conceptually simple, equally suitable for metallic and gapped systems, and yields good numerical accuracy for a range of 1s and 2p binding energies.

\emph{Method}.---Experimental core electron binding energies in solids are typically reported relative to the Fermi level. In contrast, two different conventions have been used for reporting calculated core electron binding energies in solids. In references \cite{ozaki_absolute_2017} and \cite{walter_offset-corrected_2016}, where the $\Delta$SCF method was used, theoretical binding energies were reported relative to the Fermi level, whereas in \cite{aoki_accurate_2018} and \cite{zhu_all-electron_2021} where the GW method was used, theoretical values were reported relative to the energy of the highest occupied state. These two definitions are equivalent for metals, but in the latter case, for gapped systems the point of reference is the valence band maximum  (VBM). The latter convention is clearly more appropriate.  In calculations of insulating solids without defects, the formal Fermi level is typically set to be right in the middle of the gap. In real materials, however, the position of the Fermi level is determined by defects or impurities in the sample, and it can lie anywhere from the VBM to the conduction band minimum (CBM). Hence, the assumption that the Fermi level lies in the middle of the band gap can lead to very large errors (up to a few eV) in calculated binding energies. If the VBM is used as the point of reference instead, then comparisons between theory and experiment are valid regardless of the position of the Fermi level relative to the band edges in the experimental sample.

The binding energy of a core electron referenced to the energy of the highest occupied state could be obtained from the difference of the results of two $\Delta$SCF calculations ($\Delta\Delta$SCF): one for the ionization of a core electron, and one for the ionization of a valence electron. A key observation is that the energy of the ground state cancels out, as shown in Eqn. 1, and the core electron binding energy is obtained as the total energy difference between two final states: the core hole state, and the lowest energy state of the $N-1$ electron system.

\begin{widetext}
\begin{equation}
E_{B} = (E_{N,ground} - E_{N-1,ch}) - (E_{N,ground} - E_{N-1,ground}) = E_{N-1,ground} - E_{N-1,ch},
\label{eqn1}
\end{equation}
\end{widetext}

where $E_{N,ground}$, $E_{N-1,ground}$, and $E_{N-1,ch}$ denote the total energies of the ground state of the $N$ electron system, the ground state of the $N-1$ electron system, and the core hole state, respectively, and $E_{B}$ is the calculated core electron binding energy. All core electron binding energies reported in this work are calculated using Eqn. 1.

\emph{Compuational details}.---All DFT calculations have been performed using the all-electron electronic structure program FHI-aims \cite{blum_ab_2009,knuth_all-electron_2015,havu_efficient_2009}, in which Kohn-Sham eigenstates are expanded in terms of atom-centred basis functions defined on a numerical grid. We have used the SCAN exchange correlation functional \cite{sun_strongly_2015} implemented via \textit{dfauto}\cite{strange_automatic_2001}, and relativistic effects have been accounted for using the scaled Zeroth Order Regular Approximation (scaled ZORA) \cite{van_lenthe_relativistic_1994,faas_zora_1995}. All total energies have been evaluated at the relaxed geometry of the system in the electronic ground state. The serial LAPACK eigensolver as implemented in the ELSI (Electronic Structure Infrastructure) interface was used for solving the generalized eigenvalue problem \cite{yu_elsi_2018,yu_elsi_2020}. The direct inversion of the iterative subspace (DIIS) method was used for updating the Kohn-Sham orbitals between successive SCF iterations, with a mixing parameter of 0.2 for insulating systems and 0.05 for metallic systems \cite{pulay_convergence_1980,pulay_improvedscf_1982}. Using these settings, calculations with a core hole in a 2p orbital sometimes failed to converge: in those cases, a simple linear mixer with a mixing parameter of 0.15 was used instead. The Kerker preconditioner \cite{kerker_efficient_1981} which is enabled by default in FHI-aims was explicitly turned off in all calculations. Full details of the basis sets and the k-point sampling used in each individual calculation are given in the Supplementary Information. 

\emph{Calculating the energy of the core hole state}.---In order to calculate $E_{N-1,ch}$, it is necessary to allow all remaining core and valence electrons to fully relax in the presence of a spin-polarized, localized core hole. The issue of localization of the core hole requires special attention. In ground state DFT, core electrons are delocalized over all symmetry-equivalent atoms. However, in order to obtain binding energies that are comparable to experimental values, in the calculation of the final state with a core hole, the core orbital must be localized by manually breaking the symmetry. This has been discussed in the context of polyatomic molecules in references \cite{bagus_localized_1972,klein_nuts_2021}. The authors have previously proposed a procedure for creating a localized core hole in polyatomic molecules \cite{kahk_core_2018,kahk_accurate_2019}. In this study, a slightly modified version of this procedure is used. First, in order to localize one of the core eigenstates at a particular atomic site, a fictitious extra nuclear charge of 0.1 $e$ is added to the target atom, and the self consistent field is allowed to converge in the presence of this fictitious charge. Net neutrality is enforced by adding a compensating uniform background charge. Next, the occupancy of the localized core eigenstate is set to zero in one of the spin channels, the fictitious extra nuclear charge is removed, and using the previously obtained Kohn-Sham eigenstates as the initial guess, the self-consistent field is converged again. Again, a compensating uniform background charge is introduced to keep the system neutral. Between SCF iterations, the Maximum Overlap Method (MOM) is used to keep track of the localized core eigenstate, in case the energy ordering of the core orbitals changes \cite{gilbert_self-consistent_2008}. We emphasize that in the final calculation, there are no fictitious added point charges, and the only constraint that is applied is the non-Aufbau-principle occupation of the Kohn-Sham eigenstates. We have found this procedure for creating a localized core hole to be highly robust for all materials and core levels that we have considered.

\begin{figure}
	\centering
	\includegraphics[width=2.0in]{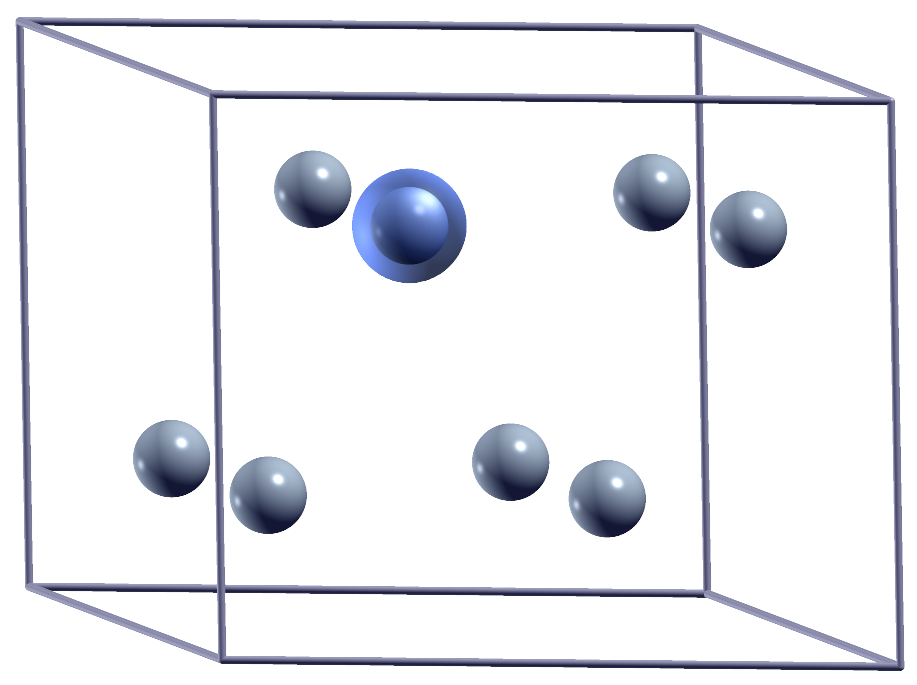}
	\caption{A localized core hole in a $2\times2\times1$ supercell of magnesium. An isosurface of the probability density of the empty core eigenstate (isovalue $= 1\times10^{-6}$) in the converged calculation of $E_{N-1,ch}$ is shown in blue. The positions of the Mg atoms are shown in gray.}
	\label{Fig_ch_isosurface}
\end{figure}

The successful localization of a core hole can be verified by examining the spatial distribution of the unoccupied core eigenstate in the converged calculation. As an illustrative example, we consider a Mg 1s core hole in a $2\times2\times1$ supercell of magnesium metal. An isosurface plot of the vacant core eigenstate is shown in Figure \ref{Fig_ch_isosurface}. The empty core orbital is found to be localized at one of the magnesium atoms and its wavefunction has almost perfect spherical symmetry. We have also examined the electronic structure of the core hole final state using Mulliken analysis \cite{mulliken_electronic_1955}. The results are given in Table \ref{Table_mulliken}. The per-atom charge analysis indicates that the magnesium atom with a core hole carries a positive charge that is somewhat less than unity, and the rest of the positive charge is distributed among the remaining Mg atoms. A closer look at the populations of the different angular momentum channels indicates that in fact, the occupancy of basis functions of p-symmetry is higher for the atom with a core hole. This is indicative of screening of the core hole by the sea of conduction electrons. The per-atom spin analysis indicates that almost the entire spin of the total system is carried by the atom with a core hole, and the spin polarization arises from s states. This is consistent with the removal of one electron from a localized s-orbital. We note that the precise values of the Mulliken populations are dependent on the chosen basis set: they provide a reasonable qualitative description of the electronic structure of the core hole state, but they cannot be interpreted as a quantitative measure of the extent of screening.

\begin{table*}
	\caption{Mulliken analysis of the electronic structure of a Mg $2\times2\times1$ supercell with a localized Mg 1s core hole.}
	\begin{ruledtabular}
	\begin{tabular}{ c c c c c c c c c c }
		 &  &  \multicolumn{4}{c}{Per-atom charge analysis}  &  \multicolumn{4}{c}{Per-atom spin analysis} \\
		Atom & Electrons & Total & $s$ ($l=0$) & $p$ ($l=1$) & $d$ ($l=2$) & Total & $s$ ($l=0$) & $p$ ($l=1$) & $d$ ($l=2$)\\ 
		\hline 
		1 & 11.28 & 0.72 & 3.69 & 7.09 & 0.45 & -1.13 & -1.00 & -0.08 & 0.00 \\
		2 & 11.96 & 0.04 & 4.77 & 6.78 & 0.37 & 0.01 & 0.01 & 0.00 & 0.00 \\
		3 & 11.98 & 0.02 & 4.79 & 6.78 & 0.37 & 0.01 & 0.01 & 0.00 & 0.00 \\
		4 & 11.96 & 0.04 & 4.77 & 6.78 & 0.37 & 0.01 & 0.01 & 0.00 & 0.00 \\
		5 & 11.98 & 0.02 & 4.79 & 6.78 & 0.37 & 0.01 & 0.01 & 0.00 & 0.00 \\
		6 & 11.96 & 0.04 & 4.77 & 6.78 & 0.37 & 0.01 & 0.01 & 0.00 & 0.00 \\
		7 & 11.98 & 0.02 & 4.79 & 6.78 & 0.37 & 0.03 & 0.03 & 0.01 & 0.00 \\
		8 & 11.91 & 0.09 & 4.75 & 6.81 & 0.31 & 0.00 & 0.00 & 0.00 & 0.00 \\
		\end{tabular} 	
	\end{ruledtabular}
		\label{Table_mulliken}
\end{table*}

\emph{Calculating $E_{N-1,ground}$}.---Evaluating the total energy of the ground state of the $N-1$ electron system, $E_{N-1,ground}$, is comparatively straightforward - one electron is removed from the top of the valence band, a uniform compensating background charge is added, and the SCF is converged via standard methods without any constraints. One possible ambiguity arises with regards to the appropriate choice of the spin state for the $N-1$ electron system. When a single electron is removed from a closed-shell system by photoemission, an open-shell system in a doublet spin state is necessarily produced. However, if a fixed spin moment is enforced in a periodic calculation of a solid, the simulated system will have one extra electron in one of the spin channels per supercell, not one extra electron in one of the spin channels per a macroscopic piece of material. From a series of numerical tests, we have found that the energy difference between the lowest energy spin-unpolarized singlet state of the $N-1$ electron system and the lowest energy spin-polarized doublet state of the $N-1$ electron system vanishes as increasingly large supercells are considered. This can be rationalized by noting that the $N-1$ electron systems created by removing an electron from the highest occupied state of a supercell are metallic regardless of whether the $N$ electron system is a metal or an insulator. The use of supercells is anyway necessary to minimize the interactions between periodic images of the core hole, as discussed in the next section. Therefore, the choice of the spin state of the $N-1$ electron system has no bearing on the extrapolated core electron binding energies reported on this study, and for the sake of simplicity, spin-unpolarized calculations of the $N-1$ electron system have been performed in all cases.

\begin{figure}
	\centering
	\includegraphics[width=3.3in]{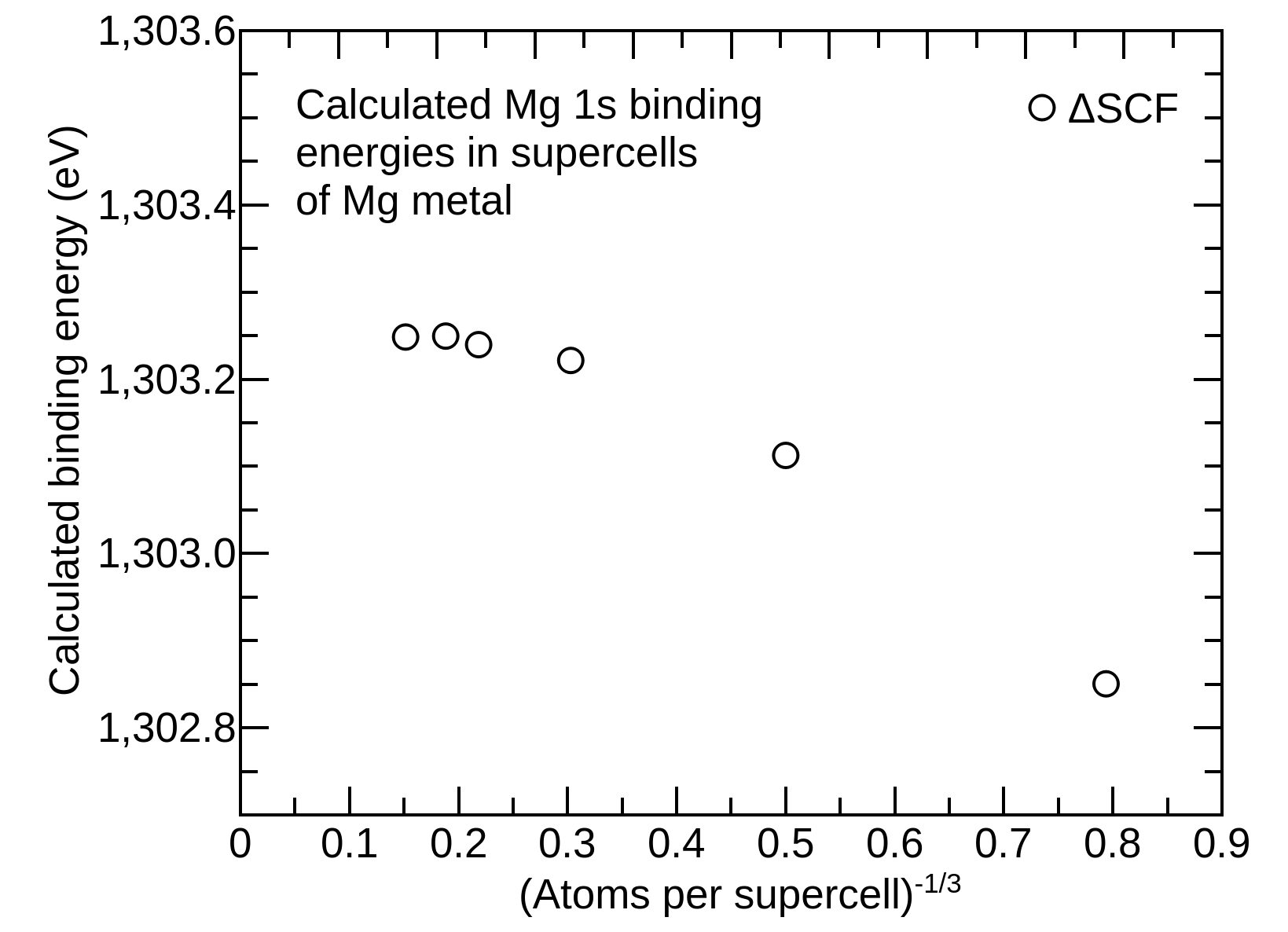}
	\caption{Dependence of the calculated Mg 1s core electron binding energy in Mg metal on the size of the supercell used in the calculation.}
	\label{Fig_convergence_Mg}
\end{figure}

\emph{Size convergence of the calculated binding energies}.---Core electron binding energies in periodic solids calculated using Eqn. 1 are affected by spurious interactions between periodic copies of the core hole and the uniform background charge. The effect of these interactions can be eliminated by performing calculations of increasingly large supercells, and extrapolating the results to the infinite size limit. The nature of the extrapolation depends on whether a core electron is removed from a metal or an insulator. In metals, interactions between periodic images of the core hole can be effectively screened by the conduction electrons. Therefore, for sufficiently large supercells, the calculated core electron binding energy converges. As a representative example of a metallic system, the dependence of the calculated Mg 1s binding energy in Mg metal on the size of the supercell is shown in Figure \ref{Fig_convergence_Mg}. Figure \ref{Fig_convergence_Mg} shows that in bulk Mg, the core electron binding energy is already converged to within 0.1 eV of the infinite size limit for a 3$\times$3$\times$2 supercell. 

\begin{figure}
	\centering
	\includegraphics[width=3.3in]{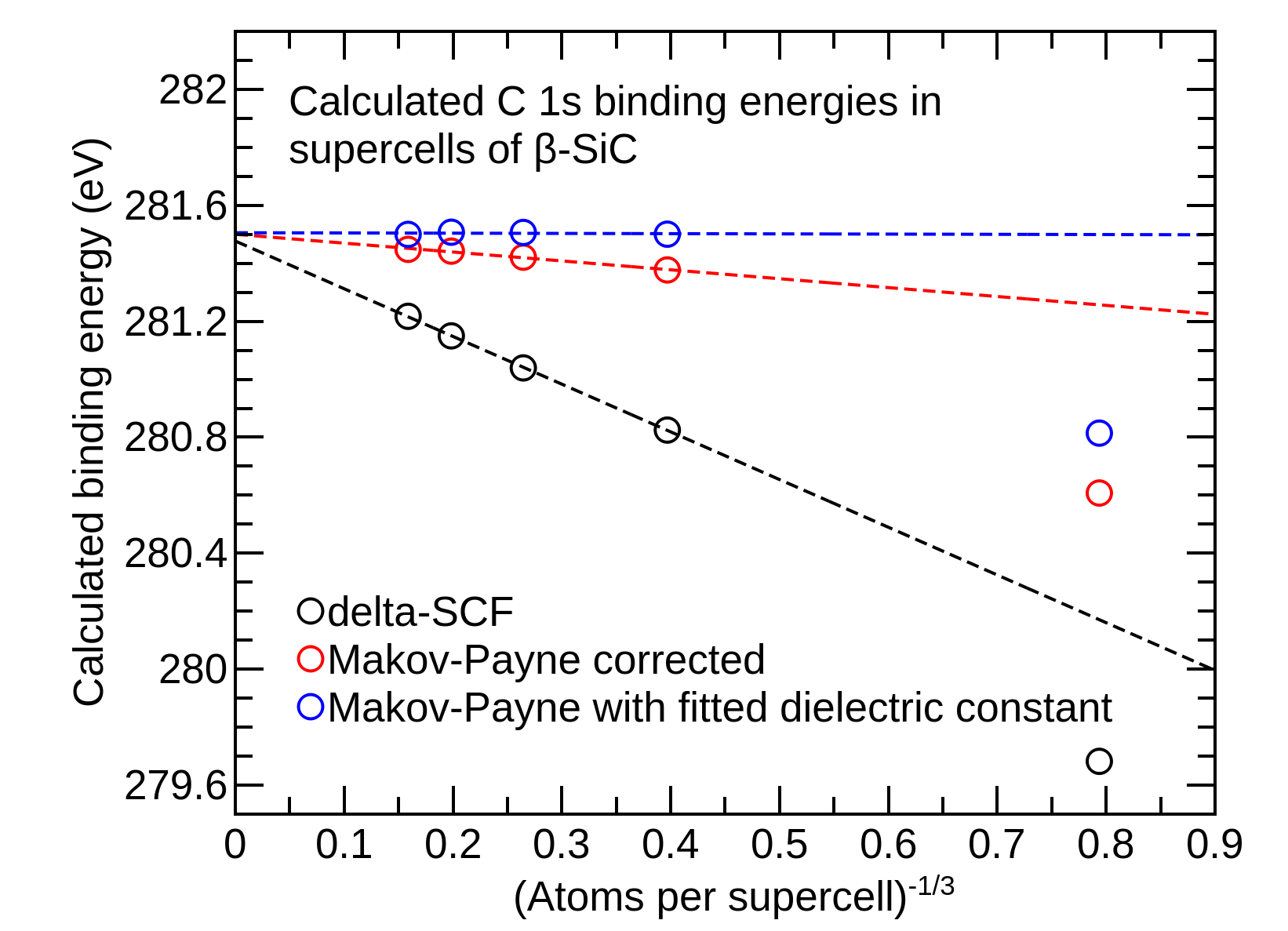}
	\caption{Dependence of the calculated C 1s core electron binding energy in $\beta$-SiC on the size of the supercell used in the calculation.}
	\label{Fig_convergence_beta_SiC}
\end{figure}

In insulators, the screened Coulomb potential of the core hole is long ranged and scales as 1/$r$ in 3-dimensional materials, where $r$ is the distance from the core hole. Absolute convergence of the calculated core electron binding energy is therefore only achieved for infinitely large supercells. Moreover, in general, it is not practically feasible to perform simulations of sufficiently large supercells to ensure that the value obtained from Eqn. 1 is within an acceptable range of the infinite size limit. However, it is observed, that for all but the smallest supercells (i.e. the unit cell, in which case periodic images of the core hole are sometimes located on adjacent atoms), plots of $E_B$ vs the inverse cube root of the number of atoms in the supercell (proportional to the inverse of the distance between two periodic copies of the core hole) yield straight lines. The extrapolated value of the core electron binding energy for an infinitely large supercell is given by the y-axis intercept of the line of best fit. As a representative example of an insulating system, the dependence of the calculated C 1s binding energy in $\beta$-SiC on the size of the supercell is shown in Figure \ref{Fig_convergence_beta_SiC}.

For insulating systems, finite size corrections akin to those used in calculations of charged defects can also be applied to calculated core electron binding energies. In this work, we examine the performance of the correction scheme due to Makov and Payne \cite{makov_periodic_1995}. In the Makov-Payne scheme, the finite size supercell correction is given by Equation \ref{eqn2},

\begin{equation}
E_{corr} = \frac {q^2 \alpha} {2 \epsilon L} - \frac {2 \pi q Q} {3 \epsilon \Omega},
\label{eqn2}
\end{equation}

where q is the defect charge state, $\alpha$ is the Madelung constant, $\epsilon$ is the dielectric constant of the material, $L$ is the lattice constant of the lattice formed by the supercells, $Q$ is the quadrupole moment of the charge induced by the defect, and $\Omega$ is the volume of the supercell. The first term is the energy of a periodically repeated point charge in a uniform neutralizing background, scaled by the bulk dielectric constant. The second term depends on the quadrupole moment of the defect charge distribution. In this work, the ``defect" is a localized core hole, and its charge distribution is approximated to that of a point charge. In this case, the second term vanishes ($Q = 0$). The value of $q^2\alpha/2L$ can be determined numerically for each supercell: this has been done in FHI-aims (sample input file given in the Supplementary Information). Then, $E_{corr}$ is obtained by dividing this value by the experimental bulk dielectric constant. Published high-frequency dielectric constants have been used, as the nuclei can be assumed to remain stationary on the time scale of core level photoemission.

The performance of the Makov-Payne correction for C 1$s$ binding energies in $\beta$-SiC is illustrated in Figure \ref{Fig_convergence_beta_SiC}. The corrected core electron binding energies are found to show a much weaker dependence on the size of the supercell, and the extrapolated value for an infinite supercell obtained from the corrected binding energies is very similar (difference $<$ 0.05 eV) to the extrapolated value of the uncorrected binding energies. 

A modified version of the Makov-Payne correction scheme has been proposed, where the ``optimal" value of the bulk dielectric constant, $\epsilon_{opt}$, is chosen such that the size-dependence of the calculated results is minimized \cite{castleton_managing_2006,hine_supercell_2009}. For the C 1s core level in $\beta$-SiC, we have found that the plot of the calculated binding energy versus the inverse of the cube root of the number of atoms per supercell becomes flat for $\epsilon_{opt} = 5.33$. The value of the calculated core electron binding energy obtained from the modified Makov-Payne scheme is again very similar to the extrapolated values from the other two methods. More advanced finite size correction schemes \cite{lany_accurate_2009,lany_assessment_2008,freysoldt_electrostatic_2011,freysoldt_first-principles_2014,freysoldt_first-principles_2018} have been developed for calculations of charged defects, where the defect charge is often spread over a larger area and the Makov-Payne correction becomes inadequate, however, the simple Makov-Payne scheme is considered adequate for the present study due to the strong localization of the core hole.

Similar size-convergence plots for all of the other calculated binding energies reported in this work are given in the Supplementary Information. In general, we have found that in metallic systems, rapid convergence of the calculated binding energy with supercell size is always observed, and in gapped systems, the three different extrapolation methods always yield a similar extrapolated value. Graphite, which is a semimetal, is found to exhibit similar size-convergence to the insulating systems. In the rest of this manuscript, only the extrapolated values of calculated core electron binding energies are discussed.

\emph{Comparison between theory and experiment}.---A comparison between the theoretical core electron binding energies from our $\Delta$SCF calculations, and experimentally reported values is presented in Table \ref{Table_theory_vs_expt}. We emphasize that for all of the gapped systems included in Table \ref{Table_theory_vs_expt}, both the theoretical and experimental values are referenced to the VBM, not the Fermi level. In most cases, excellent agreement between the theoretical and the experimental results is observed. The mean absolute error (MAE) for the 15 core electron binding energies considered in this study is 0.24 eV. The MAE for the 11 1s binding energies is 0.29 eV, and for the four 2p binding energies is 0.11 eV. On average, very small errors are obtained for metallic systems (MAE = 0.08 eV), whereas slightly larger errors (MAE = 0.38 eV) are found for insulators. However, a significant part of the MAE for insulators comes from just one material, namely BeO. In BeO, our calculations overestimate both the Be 1s and O 1s core electron binding energies by approximately 1 eV. Since both core electron binding energies are overestimated by a similar amount, we speculate that the error may be related to the calculation of the ground state of the $N-1$ electron system, which is a shared point of reference. In particular, we hypothesize that DFT with the SCAN functional may fail to accurately predict the position of the VBM in BeO due to the known limitations of semilocal exchange-correlation functionals in describing very ionic materials with large band gaps. Considerably smaller errors are observed for all other insulators with smaller band gaps, and all metallic systems.

\begin{table}
	\caption{A comparison of calculated and experimental core electron binding energies in solids.}
	\begin{ruledtabular}
	\begin{tabular}{ c c c c c c }
		 \multirow{2}{*}{Solid} & Core & Theor. &  Expt.  & \multirow{2}{*}{Ref.} & Error \\
		 & level & $E_B$ (eV) & $E_B$ (eV) &  & (eV) \\
		\hline 
		Li & Li 1s & 54.88 & 54.85 & \cite{shek_soft_1990,contour_analysis_1979,wertheim_electronic_1980,kowalczyk_x-ray_1973} & 0.03 \\
		Be & Be 1s & 111.88 & 111.85 & \cite{powell_elemental_1995} & 0.03 \\
		\multirow{2}{*}{Na} & Na 1s & 1071.56 & 1071.75 & \multirow{2}{*}{\cite{barrie_auger_1975,citrin_high-resolution_1973,kowalczyk_x-ray_1973}} & -0.19 \\
		& Na 2p & 30.65 & 30.51 & & 0.14 \\
		\multirow{2}{*}{Mg} & Mg 1s & 1303.25 & 1303.24 & \multirow{2}{*}{\cite{jennison_calculation_1984,yoshimura_degradation_2007,powell_elemental_1995,ley_many-body_1975,peng_reactions_1988,darrah_thomas_valence_1986}} & 0.01 \\
		& Mg 2p & 49.69 & 49.79 & & -0.1 \\
		Graphite & C 1s & 284.44 & 284.41 & \cite{kieser_new_1976,johansson_calibration_1973,xie_ultrahigh_1992,hamrin_valence_1970,estrade-szwarckopf_photoelectron_1992} & 0.03 \\
		\multirow{2}{*}{BeO} & Be 1s & 110.79 & 110.0 & & 0.79 \\
		& O 1s & 528.86 & 527.7 & \cite{koh_valence_2019,hamrin_valence_1970} & 1.16 \\
		\multirow{2}{*}{hex-BN} & B 1s & 188.42 & 188.35 & \multirow{2}{*}{\cite{hamrin_valence_1970,henck_direct_2017}} & 0.07 \\
		& N 1s & 396.39 & 396.0 & & 0.39 \\
		Diamond & C 1s & 284.43 & 284.04 & \cite{gaowei_annealing_2012,mcfeely_x-ray_1974,kono_electron_2014,maier_electron_2001} & 0.39 \\
		\multirow{2}{*}{$\beta$-SiC} & Si 2p & 99.24 & 99.2 & \multirow{2}{*}{\cite{bermudez_growth_1988,waldrop_formation_1990,king_valence_1999}} & 0.04 \\
		& C 1s & 281.48 & 281.55 & & -0.07 \\
		Si & Si 2p & 99.17 & 99.03 & \cite{yu_measurement_1990,puthenkovilakam_valence_2004} & 0.14 \\
		\hline 
		\multicolumn{6}{r}{Mean Absolute Error = 0.24 eV}
		\end{tabular} 	
	\end{ruledtabular}
		\label{Table_theory_vs_expt}
\end{table}

It is interesting to compare the DFT-$\Delta$SCF method for predicting core electron binding energies in solids to the GW method, that was used in references \cite{aoki_accurate_2018} and \cite{zhu_all-electron_2021}. The GW approach has been highly successful at predicting the binding energies of valence electrons, and it is known to solve the famous band gap problem in DFT. The use of the GW method for the calculation of core electron binding energies was investigated in detail in references \cite{golze_core-level_2018} and \cite{golze_accurate_2020}, and it was shown that G0W0 calculations with a PBEh45 starting point or eigenvalue-self-consistent GW calculations can yield accurate absolute core electron binding energies in molecular systems. However, these studies also found that a full frequency treatment of the self-energy is essential to obtain meaningful results, that G0W0 calculations of core electron binding energies show a very strong starting point dependence, and that extrapolation of the calculated binding energies for a series of basis sets to the complete basis set limit is required. The principal advantage of the DFT-$\Delta$SCF method over eigenvalue self-consistent full-frequency GW calculations is the significantly lower computational cost of the former theory. However, we also note that the mean absolute error found in this work (0.24 eV) is significantly lower than the mean absolute errors reported in previous GW calculations of absolute core electron binding energies in solids (0.53 eV and 0.57 eV eV in references \cite{aoki_accurate_2018} and \cite{zhu_all-electron_2021} respectively). In the view of the authors, the main motivation to go beyond $\Delta$SCF calculations for modelling core level photoemission arises when knowledge of the full spectral function including lifetime broadening and satellite peaks, instead of just the position of the fully screened (main) peak is required.

\emph{Conclusions}.---The results presented in this study highlight the versatility of the DFT-$\Delta$SCF approach for predicting absolute core electron binding energies. In particular, they establish that the same computational framework that was previously found to yield accurate absolute core electron binding energies in molecular systems also produces good results for both metallic and insulating solids. This is significant, as many experimental XPS studies are performed on samples that contain a mixture of different phases, e.g. molecules on surfaces, nanocomposites, etc. The reported calculations do not rely on any empirical parameters, and they have the same computational cost as standard ground state DFT calculations with a semilocal functional. 

We have taken advantage of the analogy between core holes and charged defects to invoke a well-established correction scheme for reducing finite size errors in calculated core electron binding energies. For the largest supercells considered for each material, the corrected binding energy lies within 0.12 eV of the extrapolated value. This suggests that for systems where the construction of supercells is not feasible, e.g. if the unit cell already contains hundreds of atoms, an accurate estimate of the absolute core electron binding energy can be obtained from a single $\Delta$SCF calculation.

\bibliography{Llibrary_v3}

\end{document}